\documentclass[prd,nofootinbib,10pt]{revtex4}
\usepackage{natbib}
\usepackage{amssymb,amsbsy,amsmath,amsfonts}
\usepackage{graphicx}% Include figure files
\usepackage{float}
\usepackage{rotating}
\usepackage{color}

\begin{document}
\title {The strangeness content of the nucleon from effective field theory and phenomenology}

\author{J. M. Alarc\'on$^{1}$}
\author{L. S. Geng$^{2}$}
\author{J. Martin Camalich$^{3}$}
\author{J. A. Oller$^{4}$}
\affiliation{
$^1$Institut f\"ur Kernphysik, Johannes Gutenberg Universit\"at, Mainz D-55099, Germany\\
$^2$Research Center for Nuclear Science and Technology \& School of Physics and Nuclear Energy Engineering, Beihang University,
Beijing 100191, China\\
$^3$Department of Physics and Astronomy, University of Sussex, BN1 9QH, Brighton, UK\\
$^4$Departamento de F\'isica. Universidad de Murcia,  E-30071 Murcia, Spain}

\begin{abstract}

We revisit the classical relation between the strangeness content of the nucleon, the pion-nucleon sigma term and the $SU(3)_F$ breaking of the baryon masses in the context of Lorentz covariant chiral perturbation theory with explicit 
decuplet-baryon resonance fields. We find that a value of the pion-nucleon sigma term of $\sim$60~MeV is not necessarily
 at odds with a small strangeness content of the nucleon, in line with the fulfillment of the OZI rule.
 Moreover, this value is indeed favored by our next-to-leading order calculation.
 We compare our results with earlier ones and discuss the convergence of the chiral series as well as 
the uncertainties of chiral approaches to the determination of the sigma terms.      

\end{abstract}

\maketitle

\section{Introduction}

We dedicate this study to the interplay between the nucleon sigma terms, $\sigma_{\pi N}$ and  $\sigma_s$,
 which are defined as
\begin{eqnarray}
&&\sigma_{\pi N}=\frac{1}{2 M_N}\langle N|\hat{m}\left(\bar{u}u+\bar{d}d\right)|N\rangle,\nonumber\\
&&\sigma_s=\frac{1}{2 M_N}\langle N|m_s\bar{s}s|N\rangle.\label{Eq:STsDefinition}
\end{eqnarray}
Here,  the up, down and strange quarks masses are indicated by $m_u$, $m_d$ and $m_s$, respectively, and $\hat{m}=(m_u+m_d)/2$. 
In the following, we restrict ourselves to the isospin limit, $m_u=m_d=\hat{m}$, with the nucleon states having the Lorentz invariant 
normalization $\langle N(\mathbf{p}',s')|N(\mathbf{p},s)\rangle=2 E_N (2\pi)^3 \delta(\mathbf{p}'-\mathbf{p})$, 
where $E_N= \sqrt{M_N^2+\mathbf{p}^2}$, $M_N$ is the nucleon mass and $s$ and $s'$ are the spin indices. 

Both $\sigma_{\pi N}$ and $\sigma_{s}$  are interesting observables and their non-vanishing values  
would clearly indicate that quark masses are not zero and  give contribution to the nucleon mass.  
More precisely, the values of these two sigma terms embody the internal scalar structure of the proton and neutron. If they are small, most of the nucleon mass
stems from the confinement of the lightest quarks in typical distances around 1~fm. Another property related to the nucleon scalar structure is the 
strangeness content of the nucleon, $y$, which is 
defined as
\begin{align}
y&=\frac{2\langle N|\bar{s}s|N\rangle}{\langle N|\bar{u}u+\bar{d}d|N\rangle}
=\frac{2\hat{m}}{m_s} \frac{\sigma_s}{\sigma_{\pi N}}.
\label{Eq:Def0y}
\end{align}
Notice that if the  OZI rule (large $N_C$ prediction) were exact then $y=0$. Besides their role in understanding the mass of the ordinary matter, 
$\sigma_{\pi N}$ and  $\sigma_s$ are also necessary with respect to theoretical speculations on the origin of dark matter particles based on supersymmetry. 
An accurate determination of the sigma terms is needed to constrain the parameter space of the underlying
supersymmetric models from the experimental bounds in direct searches of weakly interacting dark 
matter particles~\cite{Bottinoetal}.

The determination of $\sigma_{\pi N}$ is feasible from $\pi N$ scattering data due to the low-energy theorem of current 
algebra \cite{cdp} that relates the value of the isospin even $\pi N$ scattering amplitude at the Cheng-Dashen point with 
the nucleon scalar form factor \cite{Cheng:1970mx,Hohler:1982ja,Gasser:1990ce}. However, the situation is much more 
obscure for the strangeness scalar form factor of the nucleon, and then for the phenomenological 
determination of $\sigma_s$ as well as of $y$. Historically \cite{Cheng:1975wm}, the path to escape this end point is based on combining the definitions of  
Eqs.~\eqref{Eq:STsDefinition} and \eqref{Eq:Def0y} as
\begin{equation}
\sigma_{\pi N}=\frac{\sigma_0}{1-y}, \label{Eq:sigmapiN-sigma0-y}
\end{equation}
where $\sigma_0$ is the nucleon expectation value of the purely octet operator $\bar{u}u+\bar{d}d-2\bar{s}s$, 
\begin{equation}
\sigma_0=\frac{\hat{m}}{2M_N}\langle N|\bar{u}u+\bar{d}d-2\bar{s}s|N\rangle.\label{Eq:Defsigma0}
\end{equation}
The point to notice is that the latter operator is the only one in the  QCD Lagrangian
 responsible  for the hadronic mass splitting 
 within an $SU(3)$ multiplet. From the experimental values of the lightest baryon octet masses, 
$M_\Xi$, $M_\Sigma$ and $M_N$, we can then calculate approximately $\sigma_0$ 
 by making use of  $SU(3)$ flavor symmetry, with the result \cite{Cheng:1975wm}
\begin{align}
\sigma_0&=\frac{\hat{m}}{m_s-\hat{m}}\left(M_\Xi+M_\Sigma-2M_N\right)\simeq 27~\text{MeV}, 
\label{Eq:sigma0GMO}
\end{align}
where we have used $m_s/\hat{m}=26(4)$~\cite{Nakamura:2010zzi}. 

Additionally, with this value for $\sigma_0$  and by assuming the OZI rule to hold, so that $y=0$,
one obtains from Eq.~\eqref{Eq:sigmapiN-sigma0-y} the naive estimation 
 $\sigma_{\pi N}\simeq 30$~MeV, that is much smaller than its phenomenological determinations 
from $\pi N$ scattering data. For instance, Gasser {\it et al.} \cite{Gasser:1990ce} obtained the 
canonical result $\sigma_{\pi N}\simeq 45$~MeV~\cite{Gasser:1990ce} in terms of  a dispersive 
analysis of the pre-90s $\pi N$ elastic scattering data.\footnote{For a detailed exposition of the dispersive methods for obtaining $\sigma_{\pi N}$ from the
analytic continuation of the $\pi N$ scattering amplitude to the Cheng-Dashen point see Refs.~\cite{Cheng:1970mx,Hohler:1982ja,Gasser:1990ce}.} 
A partial-wave analysis including the more modern $\pi N$ database carried out by the George-Washington University group \cite{wi08}, resulted in larger values of the pion-nucleon 
sigma term, $\sigma_{\pi N}=64(8)$~MeV~\cite{Pavan:2001wz}.
Besides that, a study of $\pi N$ elastic scattering in Lorentz covariant baryon chiral perturbation theory (B$\chi$PT)
agrees with the dispersive results, which depend on the data set employed~ \cite{Alarcon:2011zs}. Additionally, it 
also reveals that modern partial-wave analyses are,
in general, more consistent with different scattering phenomenology than the older ones and lead to a
relatively large value of the sigma-term, cf. $\sigma_{\pi N}=59(7)$~MeV~\cite{Alarcon:2011zs}. 
The actual value of $\sigma_{\pi N}$ has important consequences on the strangeness content of the proton since,
according to Eq.~\eqref{Eq:sigmapiN-sigma0-y} and the result for $\sigma_0$ in Eq.~(\ref{Eq:sigma0GMO}), 
all these values for $\sigma_{\pi N}$ extracted from $\pi N$ scattering data 
 would imply a very large result for $y$.
 
Now, at this point it is important to emphasize that Eq.~(\ref{Eq:sigma0GMO}) 
is an estimate obtained at leading order in a $SU(3)_F$-breaking expansion and 
the calculation of $\sigma_0$ from this equation could be affected by 
large higher order contributions. The next-to-leading order (NLO) chiral corrections 
were first calculated by Gasser in Ref.~\cite{Gasser:1980sb}. There he obtained  
$\sigma_0=35(5)$ MeV by employing a chiral model for the meson cloud around the baryon 
which only considered contributions from the virtual octet baryons. 
Within the more evolved theoretical framework of B$\chi$PT Ref.~\cite{Borasoy:1996bx} performed 
a calculation of the baryon masses and $\sigma_0$ 
in the heavy-baryon (HB)~\cite{Jenkins:1990jv} expansion up to next-to-next-to-leading order (NNLO). 
In this work, the contributions of the decuplet-baryon resonances were not implemented explicitly but 
through resonance-saturation hypothesis 
they contributed to several of the many low-energy-constants (LECs) appearing at this order.
 All in all, they reported the value $\sigma_0=36(7)$~MeV, 
which was almost identical to the NLO result obtained by Gasser 15 years earlier. 
Later, Ref.~\cite{Borasoy:1998uu} also included the decuplet-baryon resonances within HB$\chi$PT using a cut-off regularization scheme 
and still obtained basically the same result for $\sigma_0$.  One should also notice that $\sigma_{\pi N}=45$~MeV was taken 
as input in the analyses of Ref.~\cite{Borasoy:1996bx,Borasoy:1998uu}, which had a strong influence in the results of Ref.~\cite{Borasoy:1998uu}.

 By employing $\sigma_0\simeq 35$~MeV from the calculations of Refs.~\cite{Gasser:1980sb,Borasoy:1996bx,Borasoy:1998uu}
in Eq.~\eqref{Eq:sigmapiN-sigma0-y} one obtains that $y\simeq 0.2$ and $0.4$ for $\sigma_{\pi N}\simeq 45$ MeV and $\simeq 60$~MeV, respectively.  
In particular, the latter value would imply a strangeness contribution to the mass of the nucleon of $\sim$300~MeV. Although not impossible, such a scenario with a 
strong breaking of the OZI rule is theoretically implausible, moreover after the experimental evidence pointing to a negligible strangeness 
contribution in other properties of the nucleon such as its electromagnetic structure~\cite{Ahmed:2011vp} and spin~\cite{Alekseev:2010ub}.
Thus, if one gives credit to these results and translate them 
into a small value of $y$, then the present widely accepted value for $\sigma_0$ around 35~MeV clearly discredits the relatively large values for $\sigma_{\pi N}$ favored by 
the most recent analysis of the $\pi N$ scattering data, cf. $\sigma_{\pi N}=64(8)$~MeV~\cite{Pavan:2001wz} and $\sigma_{\pi N}=59(7)$~MeV \cite{Alarcon:2011zs}.

It is our aim in this work to emphasize that the situation concerning $\sigma_0$ is not settled yet, so that 
the previous conclusion does not necessarily hold.  On one hand, the result of Gasser~\cite{Gasser:1980sb} is based on 
a model calculation of the meson cloud around the nucleon, whereas Refs.~\cite{Borasoy:1996bx,Borasoy:1998uu} 
might be afflicted by the poor convergence of the chiral series typically shown by HB in the $SU(3)_F$ theory~\cite{Geng:2008mf,MartinCamalich:2010fp}.

A suitable approach that also includes explicitly the contributions from the decuplet-baryon resonances 
is a Lorentz covariant formulation of B$\chi$PT with a consistent power-counting via the extended-on-mass-shell renormalization (EOMS) scheme~\cite{Fuchs:2003qc}. 
The relativistic corrections that results in this approach, in a way preserving the exact 
analytical properties of the Green functions, have been shown to tame the poorly convergent series of the HB expansion in baryonic observables as
important as the magnetic moments~\cite{Geng:2008mf,Geng:2009hh} or masses~\cite{MartinCamalich:2010fp,Geng:2011wq}. Moreover, once a prescription 
is taken to treat the problem of the interacting Rarita-Schwinger fields~\cite{Pascalutsa:2000kd}, this scheme is straightforwardly applicable to include the contributions 
of the decuplet-baryon resonances~\cite{Geng:2009hh}. In this work we calculate $\sigma_0$ up to NLO using Lorentz covariant B$\chi$PT renormalized in the EOMS prescription
and including explicitly the effects of the decuplet. We compare the results with those obtained in the HB expansion and  estimate systematic higher-order effects through 
a partial calculation of NNLO pieces. All together, we find the 
remarkable result that the value of $\sigma_0$ becomes larger so that the 
modern experimental determinations of $\sigma_{\pi N}\sim$ 60~MeV are then consistent with a small strangeness content 
of the nucleon, or with a small OZI rule violation. A first indication that the decuplet contributions could help to solve the strangeness puzzle concerning a relatively large $\sigma_{\pi N}$ was given by the HB calculation in Ref.~\cite{Jenkins:1991bs}. 
Indirectly, this was also the case in Ref.~\cite{Bernard:1993nj} where  very large and negative values of $\sigma_s$ were obtained when demanding $\sigma_{\pi N}=45$~MeV, 
indicating a larger $\sigma_0$.

%%%%%%%%%%%%%%%%%%%%%%%%%%%%%%%%%%%%%%%%%%%%%%%%%%%%%%%%%%%%%%%%%%%%%%%%%%%%%%%%%%%%%%%%%
\section{Calculation}

\begin{figure}[t]
\includegraphics[width=12cm]{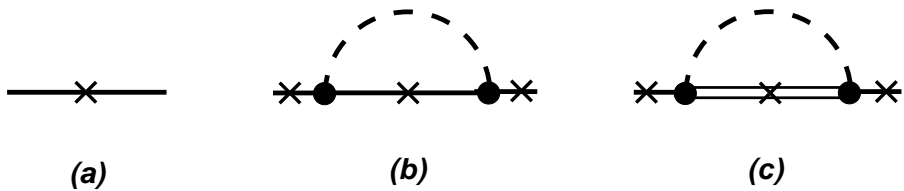}
\caption{Feynman diagrams contributing to the nucleon mass up to 
$\mathcal{O}(p^3)$ in B$\chi$PT. The internal solid lines correspond, in general, to any octet baryon, double lines 
to decuplet-baryon resonances and dashed 
lines to mesons. The black dots indicate $1^{st}$-order couplings while crosses are insertions of $\mathcal{O}(p^2)$ operators given by the LECs $b_0$, $b_D$ and $b_F$ responsible for the leading $SU(3)_F$ breaking of the baryon-octet masses.  
\label{Fig:masses}}
\end{figure}

The expressions for the sigma terms can be obtained either from the explicit calculation of the scalar form factor of the nucleon at $q^2=0$ or applying the Hellmann-Feynman theorem to the chiral expansion of its mass,
\begin{eqnarray}
\sigma_{\pi N}&=&\hat{m}\frac{\partial M_N}{\partial\hat{m}}=\frac{m_\pi^2}{2}\left(\frac{1}{m_\pi}\frac{\partial}{\partial m_\pi}+\frac{1}{2m_K}\frac{\partial}{\partial m_K}+\frac{1}{3m_\eta}\frac{\partial}{\partial m_\eta}\right)M_N+{\cal O}(p^4),\nonumber\\
\sigma_s &=&m_s\frac{\partial M_N}{\partial m_s}=(m_K^2-\frac{m_\pi^2}{2})\left(\frac{1}{2m_K}\frac{\partial}{\partial m_K}+\frac{2}{3m_\eta}\frac{\partial}{\partial m_\eta}\right)M_N+{\cal O}(p^4).\label{Eq:Hellmann-Feynman}
\end{eqnarray}
We  follow the latter strategy since the explicit expressions for the baryon masses in the different schemes treated in this paper can be directly obtained using the Appendix of Ref.~\cite{MartinCamalich:2010fp}. Thus, the chiral expansion of the sigma terms up to NLO from Eq.~\eqref{Eq:Hellmann-Feynman} is written as,
\begin{eqnarray}
&&\sigma_{\pi N}=-4(2b_0+b_D+b_F)\frac{m_\pi^2}{2}+\nonumber\\
&&\hspace{1cm}\frac{1}{(4\pi F_\phi)^2}\sum_{\phi=\pi,K,\eta}\left(\xi^{(B)}_{N,\phi} \Sigma_\pi^{(B)}(m_\phi)+
\xi^{(T)}_{N,\phi} \Sigma_\pi^{(T)}(m_\phi)\right)+{\cal O}(p^ 4),\nonumber\\
&&\sigma_{s}=-4(b_0+b_D-b_F)\left(m_K^2-\frac{m_\pi^2}{2}\right)+\nonumber\\
&&\hspace{1cm}\frac{1}{(4\pi F_\phi)^2}\sum_{\phi=\pi,K,\eta}\left(\xi^{(B)}_{N,\phi} \Sigma_s^{(B)}(m_\phi)+
\xi^{(T)}_{N,\phi} \Sigma_s^{(T)}(m_\phi)\right)+{\cal O}(p^ 4). \label{Eq:STsOp3}
\end{eqnarray}
The first line in these formulas corresponds to the LO contribution given at tree-level by the same $\mathcal{O}(p^2)$ LECs that appear in the chiral expansion of the baryon masses. 
While $b_0$ provides a $SU(3)_F$-singlet contribution that cannot be 
disentangled from the bulk mass of the octet baryons,
 the LECs $b_D$ and $b_F$ induce a splitting of octet-baryon masses  (tree-level in diagram \textbf{\textit{(a)}} in Fig.~\ref{Fig:masses}) which gives rise to the GMO relation~\cite{MartinCamalich:2010fp}. 
The second lines enclose the NLO or  $\mathcal{O}(p^3)$ corrections that stem from the loop topologies shown in Fig.~\ref{Fig:masses} \textbf{\textit{(b)}} and \textbf{\textit{(c)}}. 
Thus, the effect of virtual octet ($B$) and decuplet ($T$) baryons is explicitly accounted for. 
Their contributions are weighted by the coefficients $\xi^{(X)}_{N,\phi}$, which are combinations of $SU(3)$ Clebsch-Gordan coefficients and the meson-baryon couplings $D$, $F$ (octet contributions) and $\mathcal{C}$ (decuplet contributions). 
The loop functions $\Sigma_a^{(X)}$ depend, exclusively, on the mass of the virtual pseudoscalar meson and 
on the ones of the octet and decuplet baryons in the chiral limit, $M_B$ and $M_T$ respectively. 
Strictly speaking, the $SU(3)_F$ breaking of the baryon masses in these loops, which are represented by the crosses in Fig.~\ref{Fig:masses} \textbf{\textit{(b)}} and \textbf{\textit{(c)}}, are contributions that start at NNLO or $\mathcal{O}(p^4)$.       

For the baryon masses we use the results obtained in Ref.~\cite{MartinCamalich:2010fp} in Lorentz covariant B$\chi$PT up to $\mathcal{O}(p^3)$ in EOMS. 
The chiral loops contain divergences and analytic pieces breaking the power-counting formula~\cite{Gasser:1987rb} that are removed in dimensional regularization by the proper redefinition of the bare LECs (EOMS scheme~\cite{Fuchs:2003qc}). 
The contributions of the decuplet baryons are included taking the octet and decuplet masses in the chiral limit of approximately the same order. 
Namely, the octet and decuplet contributions are considered on the same footing for power-counting purposes and no specific expansion in $\delta=(M_T-M_B)$ is performed.
 The HB formulas~\cite{Jenkins:1991bs} can be always recovered from the renormalized covariant results by taking the non-relativistic expansion $M_B\sim M_T\sim\Lambda_{\chi SB}$.
 In particular, the HB results within the small-scale-expansion (SSE)~\cite{Hemmert:1997ye}, that it is used 
 to include explicitly 
 the decuplet resonances,\footnote{In the SSE one furthermore considers $\delta\sim p$.} are retrieved once the HB 
expansion is performed in our results~\cite{Bernard:1993nj}. 

\begin{table*}[h]
\centering
\begin{tabular}{c|c|cc|cc|}
\cline{2-6}
& &\multicolumn{2}{|c|}{ \raisebox{-1ex}[0.pt]{Octet $\mathcal{O}(p^3)$}}&\multicolumn{2}{|c|}{\raisebox{-1ex}[0.pt]{Octet+Decuplet $\mathcal{O}(p^3)$}} \\
&  \raisebox{2ex}[10.pt]{Tree level $\mathcal{O}(p^2)$}& HB & Covariant& HB-SSE& Covariant\\ 
\hline
\multicolumn{1}{|c|}{$b_D$ [GeV$^{-1}$]}&0.060(4)&0.061(4)&0.061(4)&0.315(4)&0.161(4)\\
\multicolumn{1}{|c|}{$b_F$ [GeV$^{-1}$]}&$-$0.213(2)&$-$0.502(2)&$-$0.420(2)&$-$0.704(2)&$-$0.502(2)\\
\hline
\end{tabular}
\caption{Values of the $\mathcal{O}(p^2)$ LECs $b_D$ and $b_F$ determined from the baryon octet mass splittings in the different B$\chi$PT approaches considered in this paper. \label{Table:LECs}}
\end{table*}

For the numerical values of the couplings, we use $D=0.80$ and $F=0.46$ ~\cite{Cabibbo:2003cu}. 
The decuplet coupling $\mathcal{C}$ can be fixed from the $\Delta(1232)\rightarrow\pi N$ decay rate, giving $\mathcal{C}=1.0$~\cite{Geng:2009hh}. However, there is some evidence from LQCD that this coupling is somewhat smaller~\cite{Geng:2011wq}. Indeed, an $SU(3)_F$-average among the different decuplet-to-octet pionic decay channels gives $\mathcal{C}=0.85\pm0.15$, that is the value we use.\footnote{Note that the value for $\mathcal{C}$ of the present work is different from the one often used in HB calculations~\cite{Jenkins:1990jv}. In these papers, a convention for the ``vielbein'' that is related to ours by a factor of 2 is employed. Moreover, the value we use in this paper is different to the one used in~\cite{MartinCamalich:2010fp}, explaining the slightly different decuplet results obtained here and there.} As mentioned above, the $\mathcal{O}(p^2)$ LECs $b_D$ and $b_F$ are determined using the experimental baryon-octet mass splittings. Their values for the different B$\chi$PT schemes analyzed in this paper can be found in Table~\ref{Table:LECs}~\cite{MartinCamalich:2010fp}. For the meson decay constant we also take the $SU(3)_F$-average $F_\phi\equiv1.17f_\pi$ with $f_\pi=92.4$~MeV. Variations in these values of $D$, $F$, $C$ and $F_\phi$ were 
 discussed in Ref.~\cite{MartinCamalich:2010fp} and do not influence the final results once their correlations are taken into account. 
For the masses of the pseudoscalar mesons we use $m_\pi\equiv m_{\pi^\pm}=139$~MeV, $m_K\equiv m_{K^\pm}=494$~MeV, while for the baryon masses in the loops we use the chiral-limit baryon masses obtained at LO, $M_B^{(1)}=1.151$ GeV and $M_T^{(1)}=1.382$ GeV. 
The mass of the $\eta$ meson is fixed with the Gell-Mann-Okubo mass relation, $3m_\eta^2=4m_K^2-m_\pi^2$ which is accurate enough up to the order we work.

Finally, we  restrain our analysis to $\mathcal{O}(p^3)$ despite of the fact that 
the extension of formulas to $\mathcal{O}(p^4)$ accuracy is straightforward, albeit affected by a dramatic loss of 
predictability, and have been reported in the literature~\cite{Borasoy:1996bx,Frink:2004ic,Semke:2011ez,Ren:2012aj,Ren:2013dzt}. At the latter order, 
15 new LECs contribute to the baryon masses and sigma-terms. Eight of them correspond to $\mathcal{O}(p^2)$ operators which appear through diagrams with 
the topology of a tadpole (see Ref.~\cite{Frink:2004ic} for details). These also contribute to the chiral expansion of the meson-baryon scattering amplitudes, 
although their LECs have not been determined yet from the associated experimental data or LQCD results. The other loop diagrams appearing at this order are the ones at $\mathcal{O}(p^3)$ but with the $SU(3)_F$ breaking of the baryon masses in the loop taken into account by insertions of the $\mathcal{O}(p^2)$ LECs $b_0$, $b_D$ and $b_F$ (crosses in the diagrams \textbf{\textit{(b)}} and \textbf{\textit{(c)}} of Fig.~\ref{Fig:masses}). The remaining 7 LECs correspond to $\mathcal{O}(p^4)$ operators and they renormalize the loop divergences appearing at this order. Therefore, a quantitative analysis of the sigma terms at NNLO without any further assumption on the values of the LECs (such as Large $N_c$ constraints~\cite{Semke:2011ez} or resonance saturation hypothesis estimates~\cite{Borasoy:1996bx}) is affected, at present, by a large uncertainty. On the other hand, a promising source of theoretical information on the values of the LECs is becoming available through LQCD calculations. An application in this direction within EOMS B$\chi$PT at $\mathcal{O}(p^3)$ and $\mathcal{O}(p^4)$ can be found in~\cite{MartinCamalich:2010fp} and~\cite{Ren:2012aj,Ren:2013dzt}, respectively.

Nevertheless, the analysis of part of the $\mathcal{O}(p^4)$ corrections can be useful to asses the convergence of the chiral series and to give a credible estimate on the systematic error to the $\mathcal{O}(p^3)$ results on the sigma terms due to the truncation of their chiral expansions. Indeed, we have calculated explicitly the respective corrections arising from the $SU(3)_F$ breaking of the baryon masses in the loops \textbf{\textit{(b)}} and \textbf{\textit{(c)}} of Fig.~\ref{Fig:masses}. The divergences have been renormalized in the EOMS scheme and the uncertainty on the unknown values of the $\mathcal{O}(p^4)$ LECs has been explored by varying the renormalization scale in the interval $0.7$ GeV$\leq\mu\leq1.3$ GeV. The maximal contribution obtained for these corrections in both, the octet and decuplet diagrams, is quoted as our theoretical uncertainty. That is,
\begin{eqnarray}
&&\Delta\sigma_{\pi N}^{\rm HB}\simeq20\;\,{\rm MeV}, \hspace{1cm} \Delta\sigma_{s}^{\rm HB}\simeq140\;\, {\rm MeV},\nonumber\\
&&\hspace{-0.1cm}\Delta\sigma_{\pi N}^{\rm EOMS}\simeq6\;\,{\rm MeV}, \hspace{0.8cm} \Delta\sigma_{s}^{\rm EOMS}\simeq60\;\, {\rm MeV}.\label{Eq:Systematics}
\end{eqnarray}
This explicit calculation of higher-order pieces already confirms the expectation that the convergence in the covariant approach is substantially better than the one obtained in the HB case~\cite{Geng:2008mf,MartinCamalich:2010fp}.

%%%%%%%%%%%%%%%%%%%%%%%%%%%%%%%%%%%%%%%%%%%%%%%%%%%%%%%%%%%%%%%%%%%%%%%%%%%%%%%%%%%%%%%%%%%%%%%
%>>>>>>>>>>>>>>>>>>>>>>>>>>>>>>>>>>>>>>>>>>>>>>>>>>>>>>>>>>>>>>>>>>>>>>>>>>>>>>>>>>>>>>>>>>>>>>>
%%%%%%%%%%%%%%%%%%%%%%%%%%%%%%%%%%%%%%%%%%%%%%%%%%%%%%%%%%%%%%%%%%%%%%%%%%%%%%%%%%%%%%%%%%%%%%%
\begin{table*}[h]

\centering
\begin{tabular}{c|c|cc|}
\cline{2-4}
&  \raisebox{-0.2ex}[0.pt]{ $b_0^{\rm Expt.}$ [GeV$^{-1}$]}& $y$ & $\sigma_s$ [MeV]\\ 
\hline
\multicolumn{1}{|c|}{$\sigma_{\pi N}=45(7)$ MeV} &$-$0.79(9)&$-$0.28(13)(10)&$-$150(80)(60)\\
\hline
\multicolumn{1}{|c|}{$\sigma_{\pi N}=59(7)$ MeV} &$-$0.97(9)&0.02(13)(10)&16(80)(60)\\
\hline
\end{tabular}
\caption{Value of the LEC $b_0$ and of the observables related to the strangeness of the nucleon, $y$ and $\sigma_s$, obtained in Lorentz covariant B$\chi$PT including decuplet contributions and using the phenomenological determinations of $\sigma_{\pi N}$ as input. \label{Table:strangeness}}
\end{table*}
\section{Results}

From the discussion above and the Eqs.~(\ref{Eq:STsOp3}), it is clear that the only unknown parameter in the chiral expansion of the sigma terms up to NLO is the LEC $b_0$. In the analysis that follows we calculate it by taking two phenomenological 
determinations of $\sigma_{\pi N}$, namely, $\sigma_{\pi N}\simeq45(7)$~MeV~\cite{Gasser:1990ce} and  $\sigma_{\pi N}\simeq59(7)$~MeV~\cite{Alarcon:2011zs}, and using Eq.~\eqref{Eq:STsOp3}. As a result we obtain the values for $y$ and $\sigma_s$ shown in the third and fourth columns of Table \ref{Table:strangeness}, respectively. Notice that we do not make use of LQCD results to fix our free parameters and  deliberately we only use experimental information. 

The new contributions given by the decuplet baryons are not negligible producing a $\sim$10~MeV rise on $\sigma_0$, compared 
to the situation when only the lightest octet of baryons are included in the intermediate states.
This result is very important for sigma-term physics and it has a strong impact on statements about the strangeness content of the nucleon based on the value of $\sigma_{\pi N}$, cf. Eq.~\eqref{Eq:sigmapiN-sigma0-y}.
As we can see in Table~\ref{Table:strangeness}, the determination of $\sigma_s$ 
 depends strongly on the value of the pion-nucleon sigma term due to a relative large factor $\sim m_s/(2\hat{m})\simeq13$ between the two observables.
Hence, a variation of 10~MeV in $\sigma_{\pi N}$ causes changes in $\sigma_s$ spanning more than 100~MeV. 
This factor also amplifies the uncertainty on the latter observable that propagates from the relatively small error of the former and it makes very difficult to give predictions of $\sigma_s$ with the ballpark accuracy of few tens of MeV.
 Two errors are quoted for $y$ and $\sigma_s$ in the last two columns of Table~\ref{Table:strangeness}. 
The first stems from the propagation of the uncertainty in $\sigma_{\pi N}$ and the last from the estimated ${\cal O}(p^4)$ uncertainty in our calculation, Eq.~\eqref{Eq:Systematics}.

\begin{table*}[h]

\centering
\begin{tabular}{c|c|cc|cc|}
\cline{2-6}
& &\multicolumn{2}{|c|}{ \raisebox{-1ex}[0.pt]{Octet $\mathcal{O}(p^3)$}}&\multicolumn{2}{|c|}{\raisebox{-1ex}[0.pt]{Octet+Decuplet $\mathcal{O}(p^3)$}} \\
&  \raisebox{2ex}[10.pt]{Tree level $\mathcal{O}(p^2)$}& HB & Covariant& HB-SSE& Covariant\\ 
\hline
\multicolumn{1}{|c|}{$\sigma_0$ [MeV]}&27&58(23)&46(8)&89(23)&58(8)\\
\hline
\multicolumn{1}{|c|}{$b^{\rm OZI}_0$ [GeV$^{-1}$]}&$-$0.274&$-$0.90(15)&$-$0.70(5)&$-$1.52(15)&$-$0.95(5)\\
\hline
\end{tabular}
\caption{Values  of $\sigma_0$  and the $\mathcal{O}(p^2)$ LEC $b_0$ given by the exact fulfillment of the OZI rule 
 for the different B$\chi$PT approaches considered in this paper. \label{Table:sigma0}}
\end{table*}

In order to appreciate the improvement in the chiral expansion that results by employing Lorentz covariant B$\chi$PT in the 
EOMS we compare our results for $y=0$, quite close to the last line value in Table~\ref{Table:strangeness}, with the HB$\chi$PT
 calculations with/without the decuplet-baryon resonances in Table~\ref{Table:sigma0}. 
As we can see, the corrections to the LO result on $\sigma_0$ studied are large.
 This occurs despite that the discrepancy of the Gell-Mann-Okubo  equation is correctly predicted in any of these schemes and, in fact, the description of the experimental octet mass splittings improves at $\mathcal{O}(p^3)$~\cite{MartinCamalich:2010fp}.
 As already anticipated by the calculation of the $\mathcal{O}(p^4)$ pieces in Eqs.~(\ref{Eq:Systematics}), the $SU(3)_F$ HB expansion has severe problems of convergence in the description of the sigma terms at ${\cal O}(p^3)$. 
The huge central value and errors of $\sigma_0$ for the HB-SSE expansion has to be regarded as a clear manifestation of these problems. Another one is the large variation in the value of $\sigma_0$ between HB and HB-SSE. 
On the contrary, for the covariant calculation  the difference between the calculations excluding/including the explicit decuplet of baryon resonances is only of around 10~MeV, much smaller than the difference between the LO and NLO results in the purely octet formulation of the theory.
 This indicates a stabilization of the final outcome at the ${\cal O}(p^3)$ value for the covariant case.  These conclusions are consistent with those derived from the analyses of other observables~\cite{Geng:2008mf,Geng:2009hh,MartinCamalich:2010fp} that also indicated similar problems of convergence for the HB studies in the $SU(3)_F$ sector. 
 Similar comments can be done concerning the value of $b_0$ and its variations when comparing with the different 
levels of sophistication in the calculation.
 E.g. one observes a change in $b_0$ between HB and HB-SSE in Table~\ref{Table:sigma0} that is 
a factor 3 times larger than for the covariant calculations.

 The main result of our study is having shown that $\sigma_{\pi N}\sim 60$~MeV is perfectly compatible with 
a rather accurate fulfillment of the OZI rule and, hence, with a small strangeness content of the nucleon. 
In other words, there is no argument against relatively large values of $\sigma_{\pi N}$ based on the OZI rule, as $\sigma_0$ can be afflicted by important systematics as those driven by properly accounting of the relativistic corrections and the explicit inclusion of the decuplet resonances.
 Other outcome of our work is that the value of $\sigma_0$ obtained using the experimental baryon-octet mass splittings, Lorentz  covariant B$\chi$PT in EOMS up to $\mathcal{O}(p^3)$ with explicit decuplet-baryon resonances as  degrees of freedom, favors $\sigma_{\pi N}\sim 60$ MeV. However, for this result to hold higher order corrections should be under control. 
Indeed, this seems to be the case as indicated by our calculation of a sub-set of known $O(p^4)$ diagrams. 
Unfortunately, first complete studies of LQCD results at $\mathcal{O}(p^4)$ in $SU(3)$ B$\chi$PT 
\cite{Semke:2011ez,Ren:2013dzt,Ren:2012aj,Bali:2011ks} are not conclusive on 
this respect yet,  as large variations in the sigma terms are found between different strategies.  
Ref.~\cite{Semke:2011ez} concludes $\sigma_{\pi N}=32\pm 2$~MeV, $\sigma_s=22\pm 20$~MeV and $y\simeq 0.05\pm 0.04$, 
 while Ref.~\cite{Ren:2013dzt} results with $\sigma_{\pi N}=46(2)(12)$ and $\sigma_s=157(25)(68)$.
Further work in this direction, ideally including more observables, experimental data and LQCD results to tackle the large number of unknown LECs appearing at ${\cal O}(p^4)$, will be necessary to settle this question.

  We also show two results from LQCD  without employing B$\chi$PT,  Ref.~\cite{Durr:2011mp} obtained
 $\sigma_{\pi N}=39(4)(^{+18}_{-7})$~MeV, $\sigma_s=33(14)(^{+23}_{-24})$~MeV
 and $y=0.20(7)(^{+13}_{-17})$, while Ref.~\cite{alexandrou} determines $y$ with the value $y=0.135(46)$. 
 These direct LQCD calculations clearly suggest a small value for $y$ and a subsequent contribution to the nucleon mass 
due to strangeness  of around the same size as from the lightest quark masses. 
   Other calculations supplying LQCD results with different formulations of B$\chi$PT  are 
\cite{Bali:2011ks,Shanahan:2012wh,Junnarkar:2013ac}. In these studies a small value for $y$ results, compatible 
with our own determination in the last line of Table~\ref{Table:strangeness}. There is also a tendency in the LQCD 
 results favoring the phenomenological determination $\sigma_{\pi N}\simeq 45(7)$~MeV~\cite{Gasser:1990ce}, 
although within present uncertainties 
the result of Ref.~\cite{Durr:2011mp} is compatible at the level of one sigma with the larger value $\sigma_{\pi N}=59(7)$~MeV~\cite{Alarcon:2011zs}.
% Ref.~\cite{Bali:2011ks} obtains $\sigma_{\pi N}=38\pm 12$~MeV, $\sigma_s=11(13)^{+9}_{-3}$~MeV and $y\simeq 0.05(40)(30)$. 
%Ref.~\cite{Shanahan:2012wh} gets $\sigma_{\pi N}=45\pm 6\pm 12$~MeV, $\sigma_s=21\pm 6$~MeV and $y=0.04\pm 0.01$.
%Ref.~\cite{Junnarkar:2013ac} gives for $\sigma_s$ the value $\sigma_s=49\pm 10\pm 15$~MeV. 

%%%%%%%%%%%%%%%%%%%%%%%%%%%%%%%%%%%%%%%%%%%%%%%%%%%%%%%%%%%%%%%%%%%%%%%%%%%%%%%%%%%%%%%
%%%%%%%%%%%%%%%%%%%%%%%%%%%%%%%%%%%%%%%%%%%%%%%
%%%%%%%%%%%%%%%%%%%
\section{Conclusions}

In summary, we have revisited an old empirical relation between the strangeness content of the nucleon and the pion-nucleon sigma term in the context of covariant B$\chi$PT and employing only phenomenological information. 
 Earlier estimates of $\sigma_0$ made at different levels of accuracy in $\chi$PT agreed on that a small violation of the OZI rule in the nucleon requires a value of $\sigma_{\pi N}$ close to $\sim$35~MeV.
 A long-standing puzzle \cite{dominguez} has arisen from sustained experimental evidence pointing to a value of this quantity close to 60~MeV, reinforced by the values obtained from modern $\pi N$ databases. 
We have shown that the previous calculations of $\sigma_0$ are afflicted by important systematic effects, in particular those given by relativistic corrections and by the omission of the decuplet resonances. 
Once these are incorporated, we obtain a larger $\sigma_0$ so that 
a relatively large value of $\sigma_{\pi N}$ is not necessarily inconsistent with a negligible strangeness content of the 
nucleon as currently indicated by experiment and LQCD. 
In fact, our calculation at NLO in Lorentz covariant B$\chi$PT in the EOMS with explicit decuplet-baryon resonances favors 
this scenario.

\section{Acknowledgments}

This work is partially funded by the MINECO (Spain) and FEDER (EU) funds under contract FIS2011-28853-C02-01 and the grants FPA2010-17806 and Fundaci\'on S\'eneca 11871/PI/09. We also thank the financial support from  the BMBF grant 06BN411, the EU-Research Infrastructure Integrating Activity ``Study of Strongly Interacting Matter" (HadronPhysics2, grant n. 227431) under the Seventh Framework Program of EU and the Consolider-Ingenio 2010 Programme CPAN (CSD2007-00042). JMA acknowledges support by the Deutsche Forschungsgemeinschaft DFG through the Collaborative Research Center ``The Low-Energy Frontier of the Standard Model" (CRC 1044). LSG acknowledges support from the Fundamental Research Funds for the Central Universities and the National Natural Science Foundation of China (Grant Nos. 11005007 and 11035007). JMC acknowledges the STFC [grant number ST/H004661/1] for support.

\end{document}